\documentclass[aoas,nameyear,dvips]{arximspdf}
\usepackage{graphicx}

\doi{10.1214/10-AOAS394}
\volume{5}
\issue{2A}
\pubyear{2011}
\firstpage{705}
\lastpage{724}

\makeatletter
\renewcommand{\epsilon}{\varepsilon}
\makeatother

\begin{document}
\begin{frontmatter}

\title{The mortality of the Italian population: Smoothing techniques on
the Lee--Carter model}
\runtitle{Smoothing techniques on
the Lee--Carter model}

\begin{aug}
\author[a]{\fnms{Valeria} \snm{D'Amato}\ead[label=e1]{vdamato@unisa.it}},
\author[b]{\fnms{Gabriella} \snm{Piscopo}\corref{}\thanksref{t2}\ead[label=e2]{gabriella.piscopo@unina.it}}
\and
\author[a]{\fnms{Maria} \snm{Russolillo}\ead[label=e3]{mrussolillo@unisa.it}}
\thankstext[*]{t2}{Corresponding author.}
\runauthor{V. D'Amato, G. Piscopo and M. Russolillo}
\affiliation{University of Salerno, University of Napoli Federico II
and~University~of~Salerno}
\address[a]{V. D'Amato\\
M. Russolillo\\
Department of Economics\\
\quad and Statistics\\
  University of Salerno\\
  Campus Fisciano
  84084\\ Italy\\
\printead{e1}\\
\phantom{E-mail: }\printead*{e3}} 
\address[b]{G. Piscopo\\
Department of Mathematics\\
\quad and Statistics\\
University of Napoli Federico II\\
80126 Napoli \\
Italy\\
\printead{e2}}
\end{aug}

\received{\smonth{9} \syear{2009}}
\revised{\smonth{7} \syear{2010}}

%
\begin{abstract}
Several approaches have been developed for forecasting mortality using
the stochastic model. In particular, the Lee--Carter model has become
widely used and there have been various extensions and modifications
proposed to attain a broader interpretation and to capture the main
features of the dynamics of the mortality intensity. Hyndman--Ullah
show a particular version of the Lee--Carter methodology, the so-called
Functional Demographic Model, which is one of the most accurate
approaches as regards some mortality data, particularly for longer
forecast horizons where the benefit of a damped trend forecast is
greater. The paper objective is properly to single out the most
suitable model between the basic Lee--Carter and the Functional
Demographic Model to the Italian mortality data. A comparative
assessment is made and the empirical results are presented using a
range of graphical analyses.
\end{abstract}

%
\begin{keyword}
\kwd{Lee--Carter model}
\kwd{functional demographic model}
\kwd{forecasting}.
\end{keyword}

\end{frontmatter}

\section{Introduction}
In the 20th century, the human mortality has declined globally. Such
trends in mortality reduction present risk for insurers which have
planned on the basis of historical mortality tables that do not take
these trends into account. In this regard, from the life insurance
business risk profile point of view, different risk sources have to be
evaluated. In particular, life insurance companies and private pension
managers deal with the demographic risk, which can be split in two
components: the insurance risk and the longevity risk. The insurance
risk arises from accidental deviations of the number of the deaths from
its expected values, and it is a
pooling risk, that is, it can be mitigated by increasing the number of
policies.

The longevity risk derives from improvements in the mortality trend,
which determine systematic deviations of the number of the deaths from
its expected values. These changes clearly affect pricing and reserve
allocation for life annuities and represent one of the major threats to
a social security system that has been planned on the basis of a more
modest life expectancy. The risk is of using mortality tables that do
not take these trends into account, thus underestimating the survival
probability and determining inappropriate premiums. To face this risk,
it is necessary to build projected tables including this trend. Thus,
reasonable mortality forecasting techniques have to be used to
consistently predict the trends [Brouhns, Denuit and Vermunt
(\citeyear{BDV})]. In that respect, over the years a number of
approaches have been proposed for forecasting mortality using the
stochastic model, however, the Lee--Carter model [Lee and Carter (\citeyear{LC})]
unquestionably represents a milestone in the literature.

This methodology has become widely used and there have been various
extensions and modifications proposed to attain a broader
interpretation and to capture the main features of the dynamics of the
mortality intensity [e.g., Booth, Maindonald and Smith
(\citeyear{BMS}); Haberman and Renshaw (\citeyear{RHb}, \citeyear{RH});
 Hyndman and Ullah
(\citeyear{Hu}); Renshaw and Haberman~(\citeyear{RHa}, \citeyear{RHc})].

\begin{figure}

\includegraphics{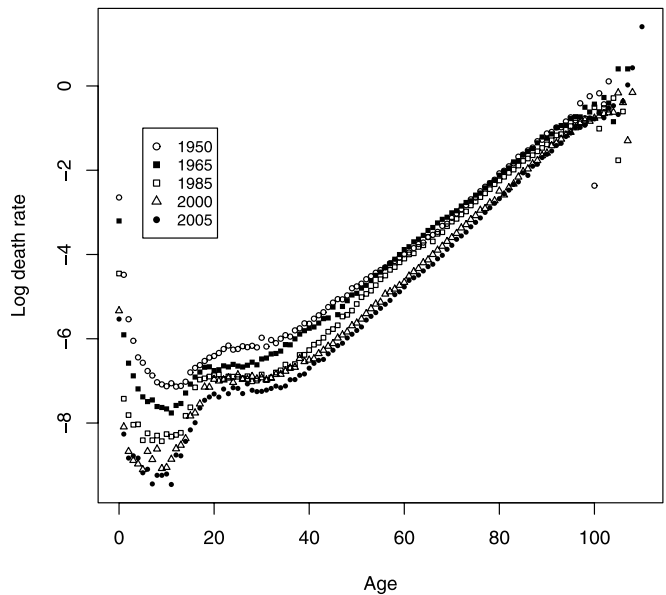}

\caption{Italian male death rates, 1950--1965--1985--2000--2005.}\label{fig1}
\end{figure}

\begin{figure}

\includegraphics{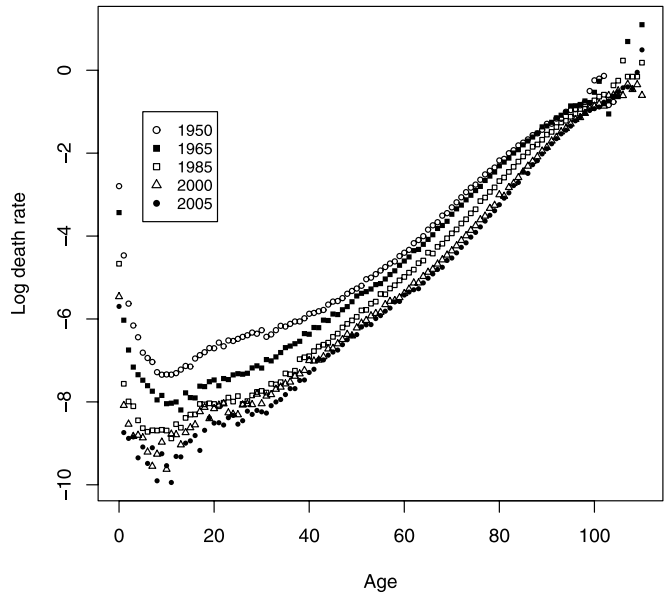}

\caption{Italian female death rates, 1950--1965--1985--2000--2005.}\label{fig2}
\end{figure}

The main statistical tool of LC is least-squares estimation via
singular value decomposition of the matrix of the log age-specific
observed death rates. In fact, the mortality data (death counts and
exposures-to-risk) have to fill a rectangular matrix. Henceforth, we
will denote with $m_{x,t}$ the observed death rates at age $x$ during
calendar year $t$, obtained by the ratio between the number of deaths,
$D_{x,t}$, recorded at age $x$ during year $t$, from an
exposure-to-risk $E_{x,t}$, that is, the number of person years from
which $D_{x,t}$ occurred. As regards the Italian population data set on
the basis of the death rates, classified by gender and individual year
from $0$ to $100$, plots of fitted values for such models suggest that
smoothing is appropriate (see Figures \ref{fig1} and \ref{fig2}). If we
look at Figures \ref{fig1} and \ref{fig2}, we can notice the random
variations in the data, especially for ages between 0 and 10, where the
reductions in the death rates are stronger. Moreover, we can notice
also for older ages the irregularities are pronounced. These
irregularities in fact propagate to the life insurance premiums as well
as reserves that have to be held by insurance companies to make them
able to pay the future contractual benefit. Consequently, the model
fitting on a population has to smooth the random variations in the
data, because otherwise the resulting death rates become less
reliable.

The aim of the paper is properly to single out the most suitable model
between the basic Lee--Carter (from herein LC) and a variant of this
model, the so-called Functional Demographic Model (from herein FDM) by
Hyndman--Ullah [Hyndman and Ullah (\citeyear{Hu})], to the Italian population
demographic trend. In particular, considering the random variations in
the data, we can get an extremely accurate fit by using appropriate
smoothing techniques.
The paper is organized as follows: in Section \ref{sec2} we describe the LC
model and the FDM model; Section \ref{sec3} shows the traditional P-splines
approach for smoothing; in Section \ref{sec4} a comparative assessment among the
basic LC and FDM is performed to the Italian population, by gender
separately considered. Concluding remarks are provided in Section \ref{sec5}.

\section{The Lee--Carter model and the Functional Demographic
Model}\label{sec2}

The LC methodology is a milestone in the mortality projections
actuarial literature. The model describes the logarithm of the observed
mortality rate for age $x$ and year $t$, $m_{x,t}$, as the sum of an
age-specific component $\alpha_x$, that is independent of time and
another component that is the product of a~time-varying parameter
$\kappa_t$, reflecting the general level of mortality and an
age-specific component $\beta_x$, that represents how mortality at each
age varies when the general level of mortality changes:
%
\begin{equation}\label{eq1}
\ln{m_{x,t}}=\alpha_x + \beta_x \kappa_t + \epsilon_{x,t}.
\end{equation}
The $\epsilon_{x,t}$ component denotes the error term, which is assumed
to be homoschedastic and normally distributed. In other words, $\alpha
_x$ describes the general age shape of the age specific death rates
$m_{x,t}$, while $\kappa_t$ is an index that describes the variation in
the level of mortality to $t$. The $\beta_x$ coefficients describe the
tendency of mortality at age $x$ to change when the general level of
mortality $\kappa_t$ changes.

The LC model cannot be fitted by ordinary regression methods, because
there are no given regressors on the right-hand side of the equation;
thus, in order to find a least squares solution to the equation (\ref
{eq1}), we use the Singular Value Decomposition (SVD) method, as
suggested in Lee and Carter (\citeyear{LC}), assuming that the errors are homoschedastic.
The parameter uniqueness is specified by a different set of conditions
from (\ref{eq1}), namely, the sum of the $\beta_x$ coefficients is
equal to one and the sum of the $\kappa_t$ parameters is equal to zero.
To forecast mortality by using the LC model, we proceed by following
two steps. In the first step, we estimate the parameters $\alpha_x$,
$\beta_x$ and $\kappa_t$ using historical mortality data. In the second
step, the estimated time-dependent parameter $\kappa_t$ is modeled as a
stochastic process by an autoregressive integrated moving average
(ARIMA p, d, q) model, determined by the standard Box and Jenkins
methodology (identification--estimation--diagnosis) [Box and Jenkins
(\citeyear{BJ}); Hamilton (\citeyear{H})]. Finally, we extrapolate $\kappa_t$ through
the fitted ARIMA model to obtain a forecast of future death rates and
generate associated life table values.

The LC model has become widely used and there have been various
extensions and modifications proposed to attain a broader
interpretation and to capture the main features of the dynamics of the
mortality intensity. Hyndman and Ullah~(\citeyear{Hu}) show a particular version
of the LC methodology, the so-called Functional Demographic Model; they
propose a methodology to forecast age-specific mortality rates, based
on the combination of functional data analysis, nonparametric smoothing
and robust statistics. In particular, the approach under consideration
allows for smooth functions of age, is robust to outliers and provides
a modeling framework easy to fit to constraints and other information.

The modeling framework they propose is a generalization of the LC
me\-thod. Let $y_t(x)$ denote the log of the observed mortality rate for
age $x$ and year~$t$, $f_t(x)$ the underlying smooth function, $\{x_i
, y_t(x_i)\}, t=1,\dots,n , i=1,\dots, p$, the functional time series,
where
%
\begin{equation}\label{eq2}
y_t(x_i)=f_t(x_i)+\sigma_t(x_i)\epsilon_{t,i},
\end{equation}
with $\epsilon_{t,i}$ an i.i.d. standard normal random variable and
$\sigma_t(x_i)$ allowing for the amount of noise to vary with $x$. The
steps for forecasting $y_t(x)$ are\vadjust{\goodbreak} summarized as follows:
\begin{enumerate}
\item The data set is smoothed for each $t$ by applying penalized
regression splines. Using a nonparametric smoothing with constraint, we
estimate for each $t$ the functions $f_t(x)$ for $x \in[x_1,x_p]$
from $\{x_i,y_t(x_i)\}$ for $i=1,\dots, p$. We assume that $f_t(x)$ is
monotonically increasing for $x \geq c$ for some $c$ , that is
reasonable for mortality data. This constraint allows to reduce the
noise in the estimated curves at older ages.
\item The fitted curves
are decomposed by using a basis function expansion:
%
\begin{equation}\label{eq3}
f_t(x)=\mu(x)+\sum_{\kappa=1}^{K}{\beta_{t,k}\phi_k(x)}+e_t(x),
\end{equation}
where $\mu(x)$is a measure of location of $f_t(x)$, $\phi_k(x)$ is a set
of orthonormal basis functions, $\{\beta_{t,k}\}$ are the coefficients
and $e_t(x)\sim N(0,v(x))$.
\item To each coefficients
$\{\beta_{t,k}\}, k=1,\ldots,K$, univariate time series models are
fitted.
\item On the basis of the fitted time series models the
coefficients $\{\beta_{t,k}\}, k=1,\dots,K$, are forecasted for $t=n+1,
\ldots, n+h$.
\item The coefficients obtained in the previous step are
implemented to get the $f_t(x)$ as in equation (\ref{eq2}). From
(\ref{eq2}) the $y_t(x)$ are projected. In other words, the $y_t(x_i)$
can be expressed as the following formula obtained by combining
(\ref{eq2}) and (\ref{eq3}):
%
\begin{equation}\label{eq4}
y_t(x_i)=\mu(x_i)+\sum_{\kappa=1}^{K}{\beta_{t,k}\phi
_k(x_i)}+e_t(x_i)+\sigma_t(x_i)\epsilon_{t,i}.
\end{equation}
In particular, the $h$-steps ahead forecasts of $y_{n+h}(x)$ are given by
the formula~(\ref{eq5}):
%
\begin{equation}\label{eq5}
\hat{y}_{n+h}(x)=E[y_{n+h}(x)\vert
I,\Phi]=\hat{\mu}(x)+\sum_{\kappa=1}^{K}{\tilde{\beta}_{n,k,h}\hat{\phi}_k(x)},
\end{equation}
where $I=\{y_t(x_i); t=1,\dots,n; i=1,\dots,p$, is the observed data,
$\Phi$ the set of basis functions, $\tilde{\beta}_{n,k,h}$ corresponds
to the $h$-step ahead forecast of~$\beta_{n+h,k}$ having been estimated
time series $\hat{\beta}_{1,k,},\dots,\hat{\beta}_{n,k,}$.
\item
Finally, in order to determine confidence intervals for mortality
projections, the variance of error terms in (\ref{eq2}) and (\ref{eq3})
is calculated. In particular, the forecast variance is written from
(\ref{eq4}):
%
\begin{equation}\label{eq6}
\zeta_{n,h}(x)=\operatorname{Var}[y_{n+h}(x)\vert
I,\Phi]=\hat{\sigma}_{\mu}^{2}(x)+\sum_{\kappa=1}^{K}{u_{n+h,k}}\hat
{\phi}_{k}^{2}(x)+v(x)+\sigma_{t}^{2}(x),
\end{equation}
where $\hat{\sigma}_{\mu}^{2}(x)$ the variance of the smooth estimate
$\hat{\mu}(x)$ depends on the
smoothing technique, $u_{n+h,k}=\operatorname{Var}(\beta_{n+h,k}\vert\beta_{1,k},\dots
,\beta_{n,k})$ are obtained by the time series model, being $\sigma
_{t}^{2}(x)$ the variance of the $y_t(x)$ and the $v(x)$ is the model
error variance estimated by averaging $\hat{e}_{t}^{2}(x)$ for each
$x$.
The prediction interval for $y_t(x)$ is represented by the following:
\[
\hat{y}_{n,h}(x)\pm z_{\alpha}\sqrt{\zeta_{n,h}(x)}
\]
assuming normally distributed the sources of error, where $z_{\alpha}$
is the $1-\frac{\alpha}{2}$ standard normal quantile.

In order to measure the uncertainty in the mortality projections,
prediction intervals can be derived applying bootstrap techniques
[Efron and Tibshirani~(\citeyear{ET})]. They are particularly useful where
theoretical calculation with the fitted model is too complex, as in the
case where the computation of interval forecast is not straightforward
[Koissi, Shapiro and Hognas (\citeyear{KSH})].

\end{enumerate}

\section{P-splines approach for smoothing}\label{sec3}
Mortality data are often characterized by the presence of some outlier
data. In particular, in the case of older ages, the high variability
can be due to the small number of survivors in the population. This
represents a common problem when estimating mortality rates for groups
aged 90 and more. Techniques of smoothing have been implemented to
avoid this shortage of data, because the heavy variance at older ages
influences the fitting of mortality models [Delwarde, Denuit and Eilers (\citeyear{DDE})].

As suggested by Eilers and Marx (\citeyear{EM}), the
Penalized splines or P-splines is now well-established as a method of
smoothing in Generalized Linear Models. The main characteristics of the
methodology under consideration are the following:
\begin{enumerate}
\item using B-splines as the basis for the regression;
\item modifying
the log-likelihood by a difference penalty on the regression
coefficients.

\end{enumerate}
In Currie, Durban and Eilers (\citeyear{CDEa}, \citeyear{CDEb}) the mortality intensity is decomposed as
\[
\ln\mu_x(t)=\sum_{i,j}{\Theta_{i,j}B_{i,j}(x,t)}
\]
for some given 2-dimensional B-splines $B_{i,j}$ in age $x$ and
calendar time $t$, with regularly-spaced knots and $ \Theta_{i,j}$'s,
the parameters to be estimated on the basis of the data set. In order
to limit the influence of the knots on the fitted value, Eilers and
Marx (\citeyear{EM}) suggest to introduce a penality based
on finite differences of the coefficients of the adjacent B-spline;
this tecnique is called P-spline. For both age and calendar year
dimensions, the penalties have to be calculated as sums of $(
\Theta_{i,j}-2 \Theta_{i-1,j}+ \Theta_{i-2,j})^{2}$. For each of these
penalties a weight coefficient has to be selected on the basis of the
historical data set. Some authors explain that the P-splines are not so
transparent to the actuaries, especially because the choice of penalty
corresponds to a~view of the future pattern of mortality. Currie (\citeyear{C})
shows the limits of using a penalized spline to smooth the mortality
data: since the penalty depends on the parameterization, the smoothed
value is not invariant with respect to its choice. The solution
proposed by Currie is to consider a direct smoothing and replace the
penalized differences in the adjacent coefficient with the penalized
differences in the adjacent fitted values. Hyndman and Ullah (\citeyear{Hu})
prefer smoothing the data first, rather than smoothing the fitted
values, as it allows to place monotonic constraint on the smoothing
more easily.

\section{Empirical analysis}\label{sec4}

We have run the application by considering the Annual Italian male and
female mortality rates from $1950$ to $2006$ for single year of age.
The data for the Italian population are downloaded from the Human
Mortality Database and are described in the
Supplementary Material at
the end of the paper. We consider death rates for single year of age,
for ages from $0$ to $100$. For each gender and for each calendar year,
the death rates, given by the ratio between the
``Number of deaths'' and the
``Exposure to risk,'' are arranged in
a matrix for age and time.
By analyzing the changes in mortality as a function of both age $x$ and
time $t$, we have seen that mortality has shown a gradual decline over
time. To have an idea of this evolution, Figure \ref{fig1} shows the general
drop in the Italian male mortality rates during the period 1950--2005.
Improvements in mortality are not uniform across the ages and the
years: first of all, reductions in mortality rate are stronger for ages
between $0$ and $10$. As it is clear, there is an increasing variance
for higher age, especially around $X=100$.
\begin{figure}
\begin{tabular}{@{}c@{}}

\includegraphics{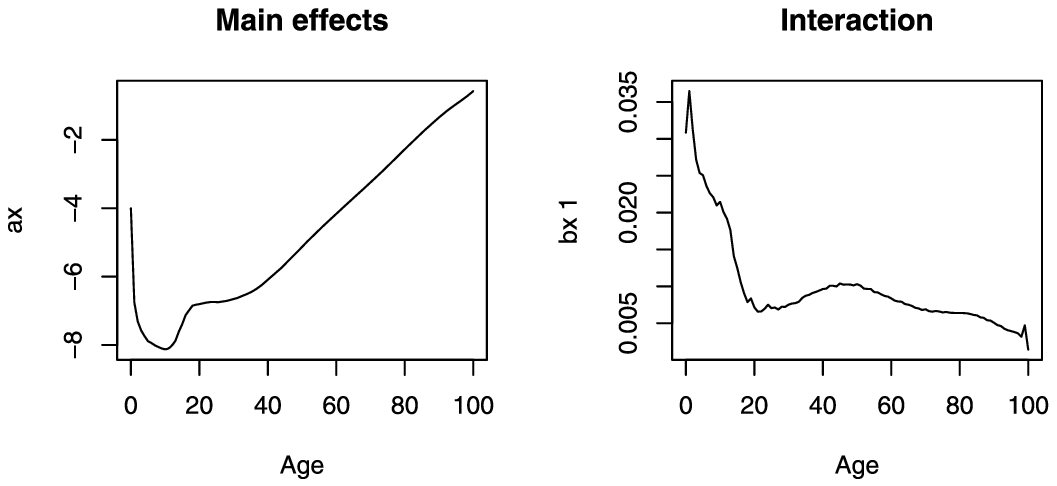}
\\[3pt]

\includegraphics{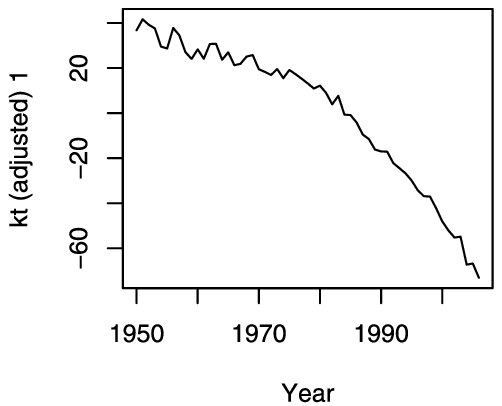}

\end{tabular}
\caption{The parameter estimates of basic Lee--Carter model on Italian
male mortality data.} \label{fig3}
\end{figure}

\begin{figure}

\includegraphics{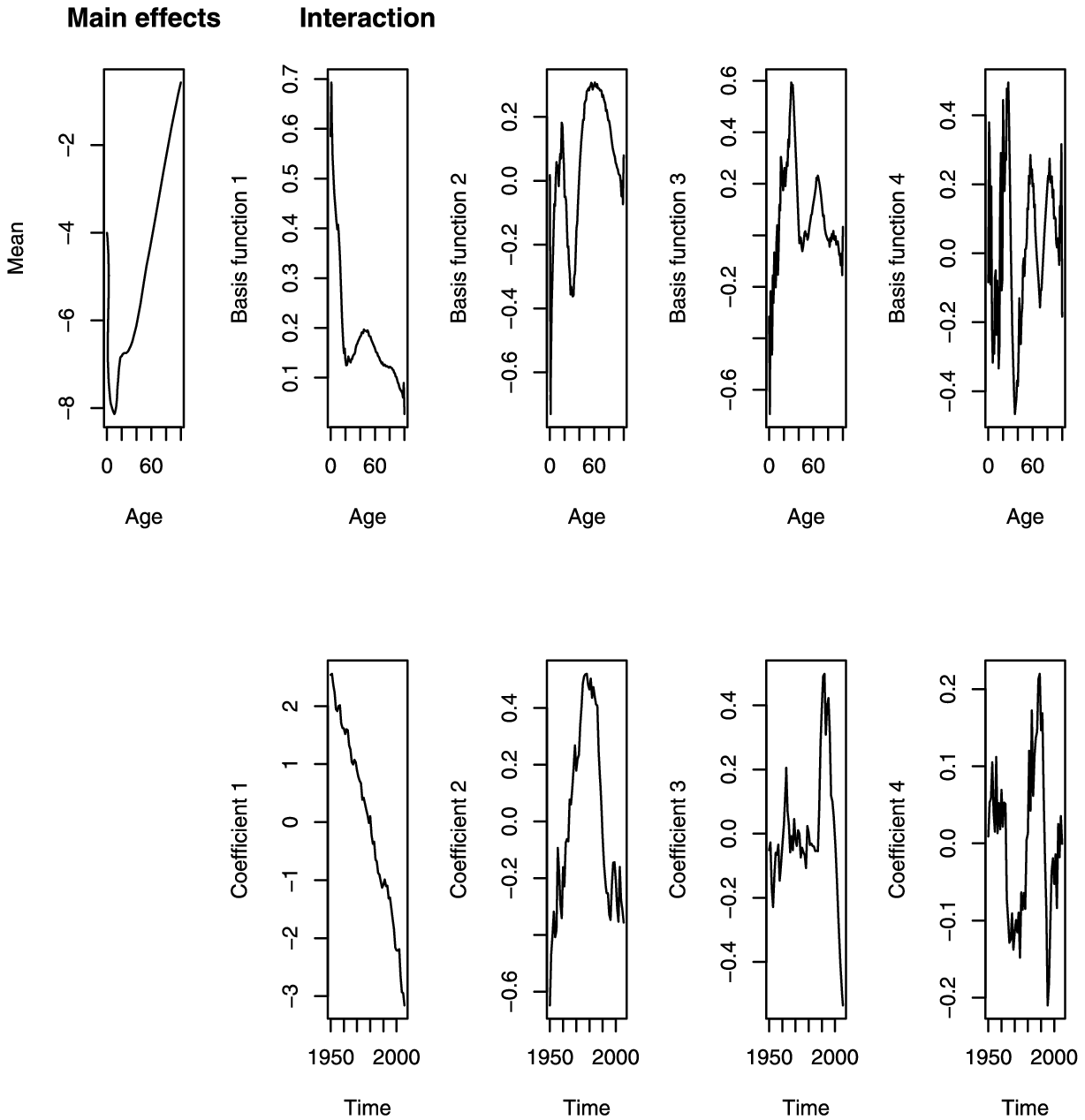}

\caption{Basis Functions of FDM and associated coefficients for Italian
male population.} \label{fig4}
\end{figure}

\begin{figure}
\begin{tabular}{@{}c@{}}

\includegraphics{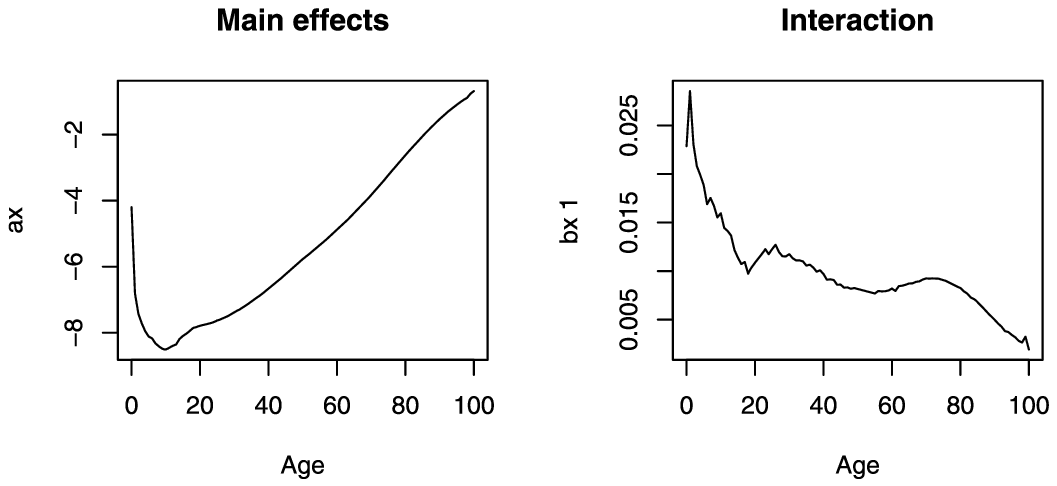}
\\[3pt]

\includegraphics{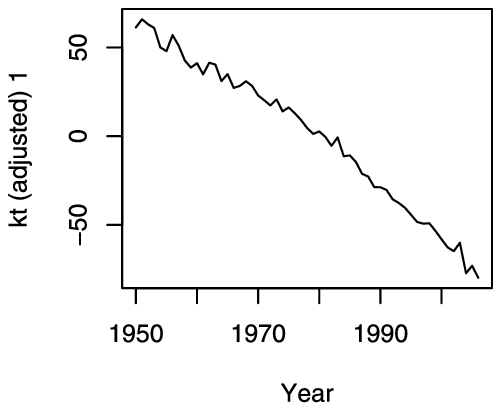}

\end{tabular}
\caption{The parameter estimates of basic Lee--Carter model on Italian
female mortality data.} \label{fig5}
\end{figure}

\begin{figure}

\includegraphics{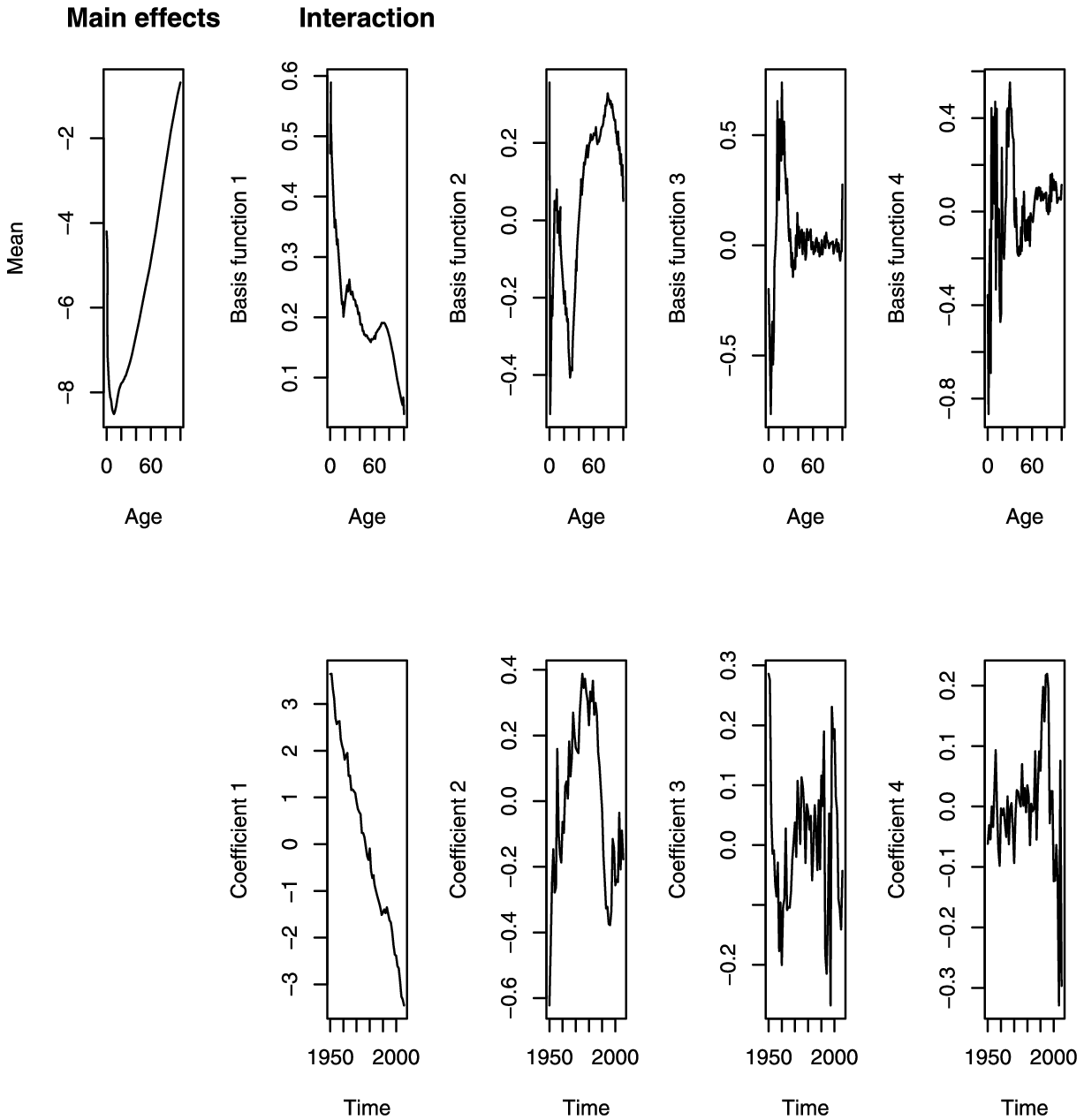}

\caption{Basis Functions of FDM and associated coefficients for Italian
female population.} \label{fig6}
\end{figure}

The first step of the application consists in fitting the basic LC
model and the FDM version to the data under consideration; Figures \ref
{fig3}--\ref{fig6} show the estimated parameters.

\begin{figure}

\includegraphics{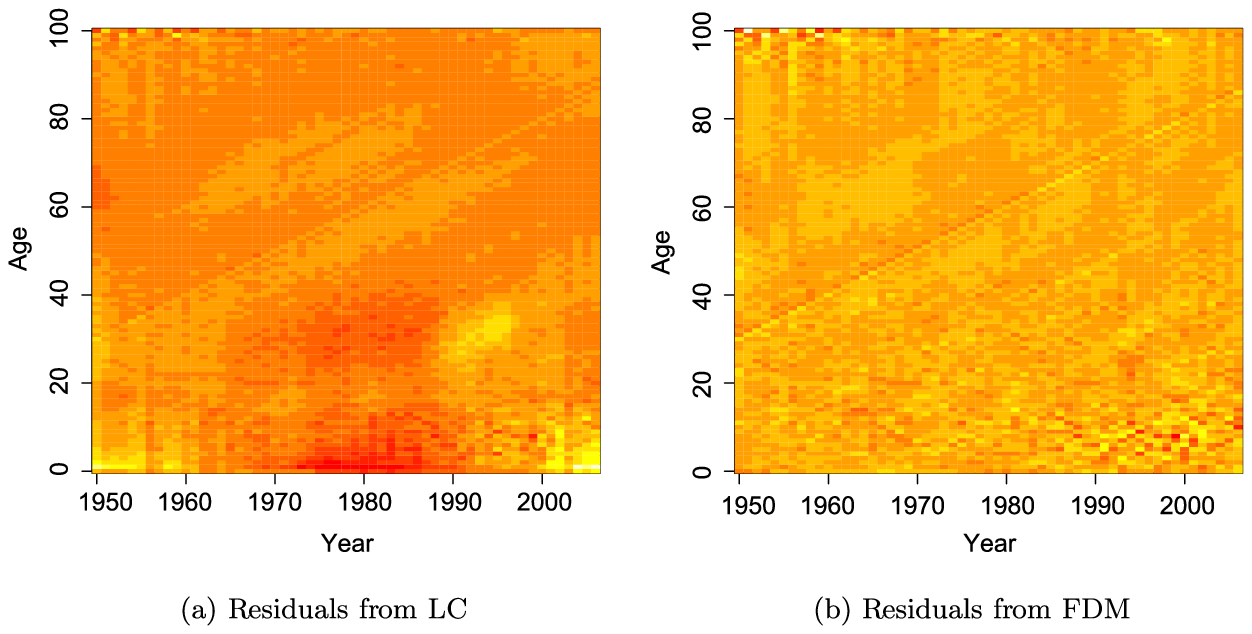}

\caption{Residuals for male data set.} \label{fig7}
\end{figure}

\begin{figure}[b]

\includegraphics{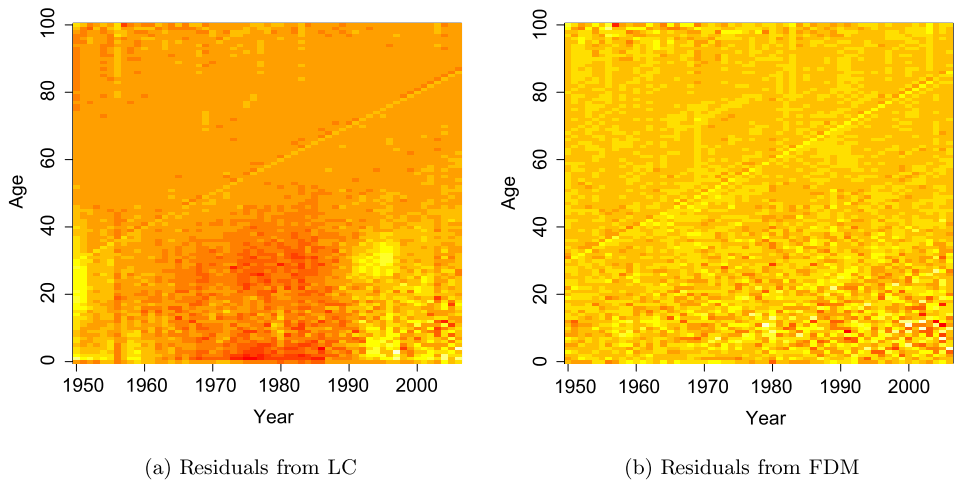}

\caption{Residuals for female data set.} \label{fig8}
\end{figure}

The percentage of variation explained by the LC for the male population
is 91.6\%, while the female is 95.7\%. This difference is due to the
features of the two data sets: as it is clear from Figures
\ref{fig1} and \ref{fig2}, male death rates show a greater dispersion at
older ages than female ones; consequently, the LC model fitted the
female data better than the male data. Shifting from the LC to the FDM,
the percentage explained by the model increases for both male and
female. In particular, if we consider the FDM model, the basis
functions explain respectively 91.8\%, 3.9\%, 1.6\%, 0.4\% of the
variation for male data and 96.0\%, 1.6\%, 0.4\%, 0.3\% of the
variation for female data. As explained in Hyndman and Ullah (\citeyear{Hu}),
the basis functions model different movements in
mortality rates across the ages. In particular, the Basis function 1
mainly models mortality changes for children. Let us have a look at the
Figures~\mbox{\ref{fig3}--\ref{fig6}}, to highlight some differences between
female and male death rates. The function fitted on female data has a
stronger slope between age $0$ and $10$ than the function fitted on
male data, so the reduction in mortality at younger ages is larger for
female than for male. Moreover, the fitted coefficient~$1$ gives us
information about the improvements in mortality at younger age. If we
look at the graphs, the mortality rates for children have dropped over
the whole period and this phenomenon is captured by the decreasing
trend of the first coefficients for both male and female, even if the
improvements are stronger for female. Following in the examination of
the movements in mortality rates modeled by the basis functions, the
Basis function 2 gives us information about the differences between
$30$ and $60$ years old, more stressed for female than for male. The
other functions are more complex and model differences between all the
cohorts; these differences are less significant for female than for
male so that the functions~$2$ and~$3$ show a lower variability between
$40$ and $100$.
We have implemented the $t$-tests on the standardized residuals for
testing the hypothesis of zero mean. The tests run for both male and
female and for FDM and LC in all cases have showed $p$-value equal to
$1$. On the contrary, the hypothesis of normality of standardized
residuals tested with the Shapiro test is always decidedly rejected;
this is a well-known limit of this mortality model [Dowd et al.
(\citeyear{DBCEM})]. However, either for male or female, there is an improvement in
the goodness of the fitting shifting from the LC to the FDM. A good fit
is achieved when the residuals are independent and identically
distributed. We have verified these conditions using contour maps (see
Figures \ref{fig7} and \ref{fig8}); moreover, we have calculated the
error measures shown in Tables \ref{table1} and~\ref{table2} as in Cairns et al. (\citeyear{CBDCE}). By comparing the
traditional LC model to the FDM one, the percentage of variation
explained by the model is higher in the FDM than in LC; in addition,
error measures are lower for both the male and female data set.

We keep running the application, in order to verify if the best fitting
of FDM depends on the smoothing on the data involved in the model or is
due to other causes.

For this reason, we smooth the data using a monotonic P-spline and then
we apply the LC method to the smoothed data; we call this procedure
LCS.
From a first analysis, we can see that the percentage variation
explained by applying the LCS model is 93.4\% for male and 97.5\% for
female. In particular, we can notice that the percentage of variance
explained increases when we shift from LC to LCS; this is not due to a
greater capacity of the model to describe the data, but to a
transformation of the same data into data less variable. Nevertheless,
the MSE of the LCS is greater than the MSE of the FDM in both data
sets.

However, it is quite possible for a model to provide a good in-sample
fit to historical data but still produce poor forecasts, that is,
forecasts that differ significantly from subsequently realized
outcomes. A good model should provide accurate fits to the historical
data as well as produce plausible forecasts. Backtesting procedures
consider what results would have been produced if the model had been
used in the past. We use this approach to test the original LC model,
the LCS and the FDM, setting out a backtesting framework that can be
used to evaluate the ex-post forecasting performance of the mortality
models. The recent literature follows this approach to evaluate the
performance of different mortality models. Lee and Miller (\citeyear{LM})
evaluated the performance of the Lee--Carter model by examining the
behavior of forecast errors comparing some error measures and producing
plots of error distributions, although they did not report any formal
test. More recently, CMI (\citeyear{CMI}) included backtesting evaluations of the
P-spline model.

In the light of this contribution, we implement a backtesting
procedure, based on the following considerations. First of all, it is
necessary to select the metric of interest, namely, the forecasted
variable that is the focus of the backtest. Possible metrics include
the mortality rate, life expectancy, future survival rates, and the
prices of annuities and other life-contingent financial instruments.
Different metrics are relevant for different purposes, for example, in
evaluating the effectiveness of a hedge of longevity or mortality risk
an important metrics could be the insurance reserves. In this paper we
focus on the mortality rate itself: our aim is to investigate the
feasibility of different mortality models rather than quantify the
impact of longevity risk on insurance product.

\begin{table}\tablewidth=270pt
\caption{Lee--Carter model}\label{table1}
\begin{tabular*}{270pt}{@{\extracolsep{\fill}}lcccc@{}}
\hline
& \textbf{ME}&
\textbf{MSE}& \textbf{MPE}& \textbf{MAPE}\\
\hline
Average across ages\\
\quad Male& $0.00786$ & $0.01947$ & $-0.01290$ & $0.03697$\\
\quad Female& $0.00190$ & $0.01462$ & \phantom{$-$}$0.00038$ & $0.01694$\\ [3pt]
Average across years\\
\quad Male& $0.78248$ & $1.88518$ & $-1.14400$ & $3.48655$\\
\quad Female& $0.18968$ & $1.40760$ & \phantom{$-$}$0.02626$ & $1.59764$\\
\hline
\end{tabular*}
\end{table}

\begin{table}[b]\tablewidth=270pt
\caption{Functional Demographic Model}\label{table2}
\begin{tabular*}{270pt}{@{\extracolsep{\fill}}lcccc@{}}
\hline
& \textbf{ME}& \textbf{MSE}& \textbf{MPE}& \textbf{MAPE}\\
\hline
Average across ages\\
\quad Male& $0.00001$ & $0.00367$ & $-0.01037$ & $0.02764$\\
\quad Female& $0.00001$ & $0.00403$ & \phantom{$-$}$0.00070$ & $0.01145$\\ [3pt]
Average across years\\
\quad Male& $0.00101$ & $0.30953$ & $-0.85270$ & $2.48752$\\
\quad Female& $0.00115$ & $0.34586$ & \phantom{$-$}$0.05956$ & $1.02922$\\
\hline
\end{tabular*}
\end{table}


\begin{figure}

\includegraphics{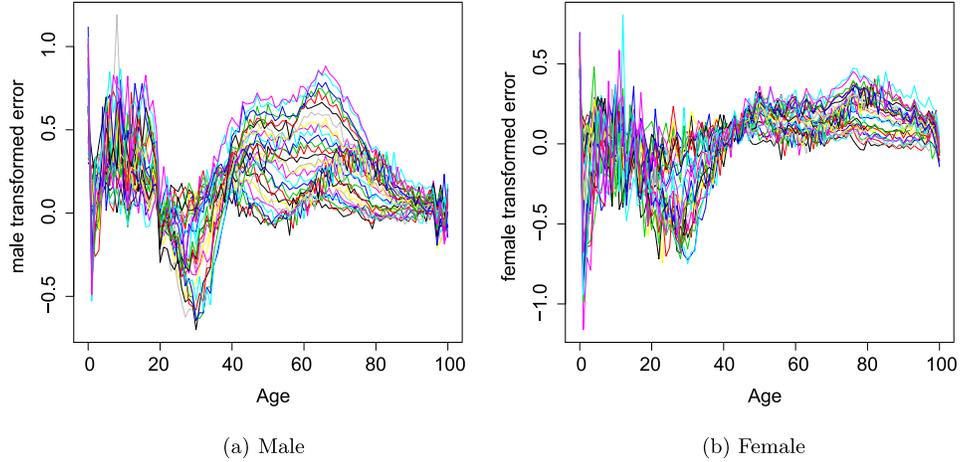}

\caption{LC Forecast errors for male and female.} \label{fig9}
\end{figure}

\begin{figure}[b]

\includegraphics{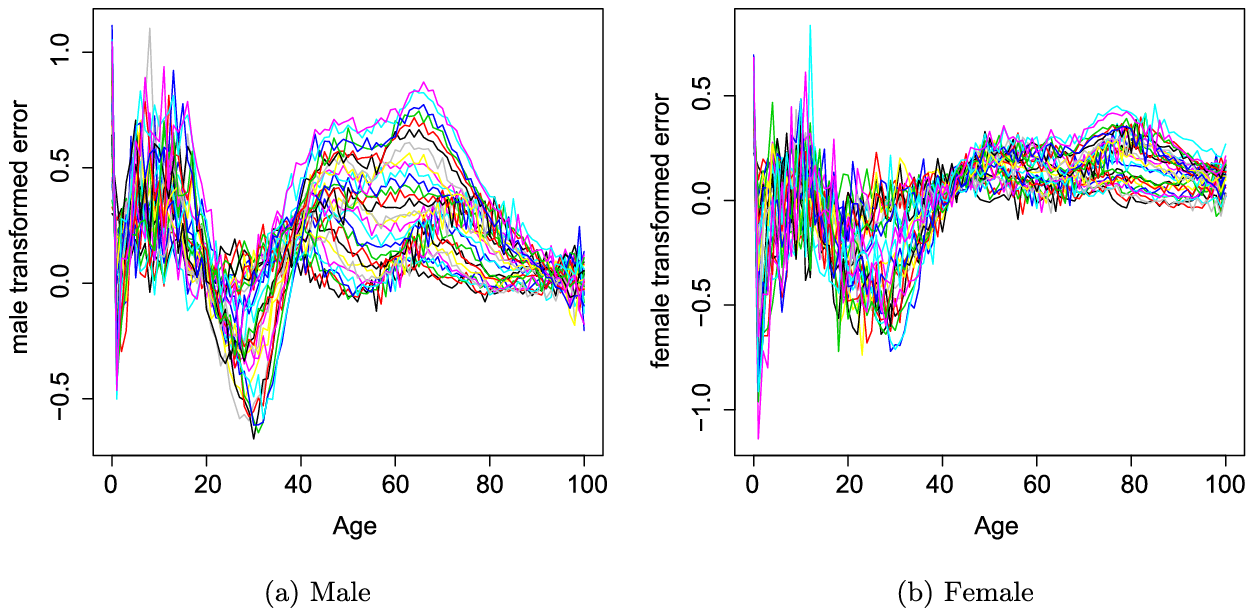}

\caption{LCS Forecast errors for male and female.} \label{fig10}
\end{figure}

\begin{figure}

\includegraphics{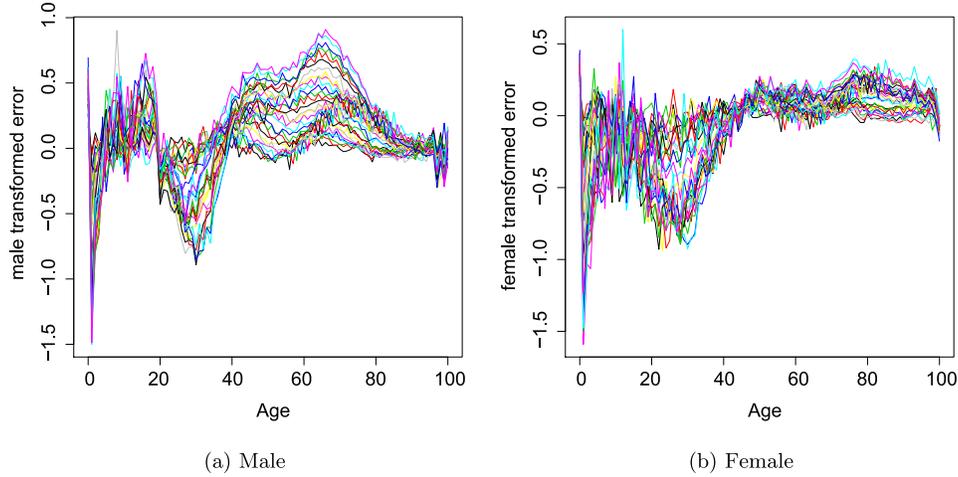}

\caption{FDM Forecast errors for male and female.} \label{fig11}
\end{figure}

Another important point is the selection of the historical
``lookback'' window and the forecast
horizon over which forecasts are made. Wang and Liu (\citeyear{WL})
highlight that as the fitted period changes, models that better perform
change. In the present paper, we focus on long-horizon forecasts,
because it is with the accuracy of these forecasts that pension plans
and life insurance companies are principally concerned. In particular,
we fit the LC, LCS and FDM from $1950$ to $1975$, thereby using a $25$
year in-sample period, then we project the mortality rates with 95\%
confidence interval from $1976$ to $2005$ according to the fitted
models and compare projections with the observed rates. It is necessary
to highlight that projections are based only on the evolution of
$k_t$: errors in $a_x$ and $b_x$ are not taken into account. In this
regard, Lee and Carter (\citeyear{LC}) found the standard errors of $a_x$ and
$b_x$ to become less significant over forecast time in comparison to the
standard error of $k_t$. Moreover, they found that by $10$ years into
the forecast of US mortality, $98$ per cent of the standard error of
life expectancy at birth was accounted for by uncertainty in $k_t$.
Figures \ref{fig9}--\ref{fig11} show the forecast error in the LC, LCS
and FDM for both male and female. In order to compare the forecast
accuracy between LC, LCS and FDM, we calculate the forecasting errors;
these are averaged over forecast years to produce mean errors indexed
by age. Moreover, we consider mean forecast errors in life expectancy
at birth averaging over forecast years.

\begin{figure}[b]

\includegraphics{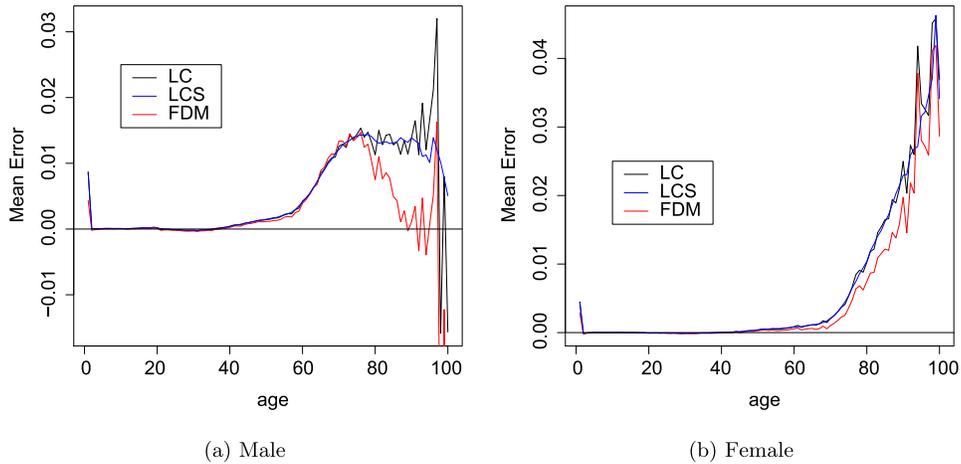}

\caption{Mean Forecast Error for male and female.} \label{fig12}
\end{figure}

Figure \ref{fig12} shows the mean forecast error by age. Shifting from
LC or LCS to FDM, a large improvement in the male forecasts accuracy in
terms of mean forecast error is obtained for ages between $75$ and
$98$; improvements also appear for the female data set, where the mean
forecast error produced by FDM is smaller than those produced by LC or
LCS almost everywhere.

\begin{figure}

\includegraphics{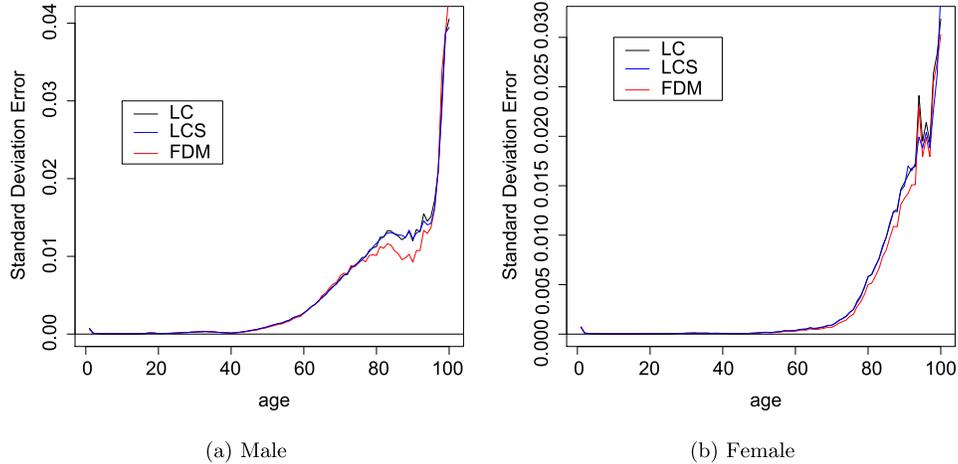}

\caption{Standard deviation of Forecast Error for male and female.}\label{fig13}
\end{figure}

\begin{figure}

\includegraphics{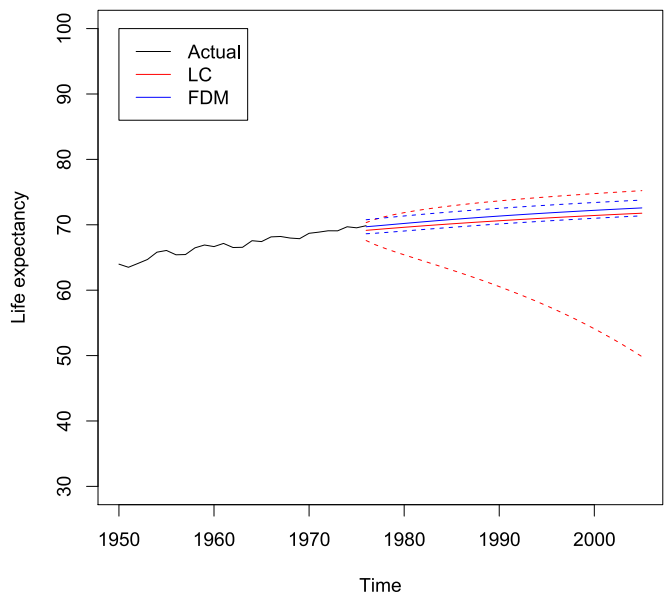}

\caption{Confidence intervals for male life expectancy.} \label{fig14}
\end{figure}

\begin{figure}

\includegraphics{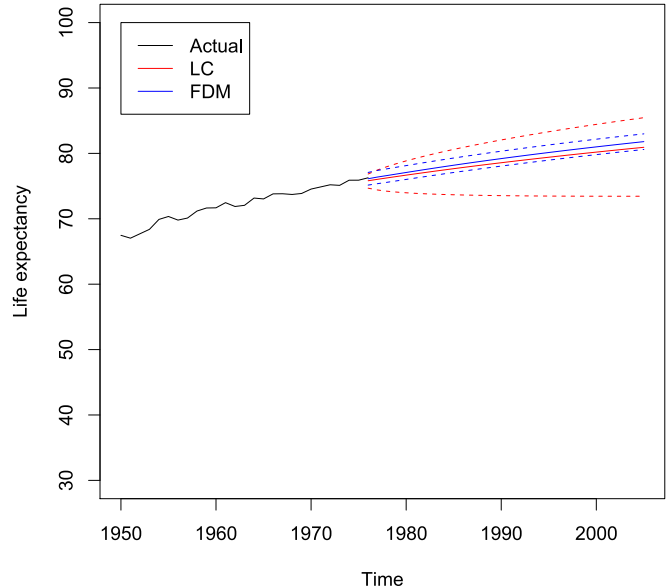}

\caption{Confidence intervals for female life expectancy.}\label{fig15}
\end{figure}

Figure \ref{fig13} shows the standard deviation of forecast error by
age: the variability in the forecasts is smaller in FDM than in LC or
LCS for both male and female almost everywhere. Tables
\ref{table3} and \ref{table4} summarize the mean and
variance of forecast error in life expectancy. The negative mean error
in life expectancy forecast means that the original LC model
underestimates the life expectancy and this results in underestimating
the capital required for cushioning against longevity risk. However,
this underestimation appears smaller for FDM for both male and female
data sets; moreover, shifting from LC or LCS to FDM, the variance of
forecast errors decreases. Finally, we derive prediction intervals as
in Lee and Carter (1992) and Hyndman and Ullah (2005). The confidence
interval is calculated at a level of 95\%. As shown in Figures
\ref{fig14} and \ref{fig15}, the interval forecasts for the life
expectancy at birth from the LC are significantly wider than those
from FDM.

\begin{table}[b]\tablewidth=255pt
\caption{Forecast error in life expectancy at birth for male data set}\label{table3}
\begin{tabular*}{255pt}{@{\extracolsep{\fill}}lccc@{}}
\hline
& \textbf{LC}& \textbf{LCS}& \textbf{FDM}\\
\hline
Mean & $-3.210096$ & $-3.215095$ & $-2.507768$\\
Variance & \phantom{$-$}$1.693628$ & \phantom{$-$}$1.691588$ & \phantom{$-$}$1.612351$\\
\hline
\end{tabular*}
\end{table}

\begin{table}\tablewidth=255pt
\caption{Forecast error in life expectancy at birth for female data set}\label{table4}
\begin{tabular*}{255pt}{@{\extracolsep{\fill}}lccc@{}}
\hline
& \textbf{LC}& \textbf{LCS}& \textbf{FDM}\\
\hline
Mean & $-1.632653$\phantom{00,} & $-1.637128$ & $-1.010379$\\
Variance & $0.6561454$ & \phantom{0,}$0.654672$ & \phantom{,}$0.49403$\\
\hline
\end{tabular*}
\end{table}

\section{Concluding remarks}\label{sec5}
In Life Insurance, primary is the relevance of the demographic
uncertainty on the portfolio liability valuations both in its
systematic and unsystematic face. The unsystematic component of the
demographic risk seems to be a risk source particularly interesting in
small portfolios, like the one at issue, for which a weak
diversification can be supposed. Unlike the risks deriving from
systematic variability, the risk due to the accidental deviations of
the number of deaths from the expected values (Mortality risk) is a
pooling risk, for which the measure becomes negligible only when the
number of contracts in the portfolio tends to infinity. The systematic
component originates from the deviations of the number of deaths from
the expected values due to the improvement in the survival trend,
taking place in the industrialized countries particularly in the last
decades. The correct capital constraint, avoiding to reserve more than
necessary, derives from the choice of the right mortality table, that
is, from the best mortality estimate. This risk source comes true in
the survival description choice and it is called Longevity risk. When
living benefits are concerned, the calculation of expected present
values (needed in pricing and reserving) requires an appropriate
mortality projection in order to avoid underestimation of future costs.
In order to protect the company from mortality improvements, actuaries
have to resort to life tables including a forecast of the future trends
of mortality (the so-called projected tables). Different approaches for
building these technical bases have been developed by actuaries and
demographers.


This paper focuses on a comparative assessment among the original LC
model and a variant of the basic methodology, the so-called FDM, for
providing accurate mortality forecasting, as regards the Italian
survival phenomenon. While it is essential to safeguard against
depicting general conclusions on the basis of individual cases, the
analysis furnishes a useful insight into the comparative performance of
the different approaches under consideration. In order to perform the
numerical analysis, we have used the demography R created by Hyndman,
Booth Tickle and Maindonald (\url{http://robjhyndman.com/software/demography/}).

The empirical results suggest the FDM framework is readily suitable to
deal with more complex forecasting problems, including forecasting of
the mortality dynamics related to extreme ages. In particular, the FDM
methodology utilizes penalized regression to smooth data using a local
algorithm which allows for contemporarily the best fitting and a fast
computational form. Although the LC is still used as a point of
reference [e.g., Renshaw and Haberman (\citeyear{RHc})], it is noted the best
performance of the FDM model.
According to our analysis applied to male and female Italian data, we
have verified that FDM data sets produce a better fitting and more
accurate forecasts than LC. We have highlighted that this improvement
is not only due to smoothing, introducing a smoothing version of the
Lee--Carter (LCS). In fact, we have obtained a better fitting and more
accurate forecast also shifting from LCS to FDM and this because FDM
explains better movements in the mortality through the basis functions
in both data sets.

The study suggests that the FDM forecast accuracy is arguably connected
to the model structure, combining functional data analysis,
nonparametric smoothing and robust statistics. In particular, the
decomposition of the fitted curve via basis functions represents the
advantage, since they capture the variability of the mortality trend,
by separating out the effects of several orthogonal components. The
empirical findings suggest the FDM framework is readily adapted to deal
with more complex forecasting problems, including forecasting of the
mortality dynamics related to extreme ages.

From the viewpoint of insurance companies, this model feature is more
desirable, because of their exposure to the variability of mortality
trend at old ages, in particular regards to post retirement
annuity-type products.
\begin{supplement}[id=suppA]
\sname{Supplement A}\label{suppA}
\stitle{Italy, Exposure to risk}
\slink[doi]{10.1214/10-AOAS394SUPPA}
\slink[url]{http://lib.stat.cmu.edu/aoas/394/supplementA.txt}
\sdatatype{.txt}
\sdescription{Italian population exposed to risk of death. The data
are downloaded from the Human Mortality database and are indexed by
calendar year during the period 1950--2005. They are divided by sex and
by single year of age for ages from 0 to~100.}
\end{supplement}
\begin{supplement}[id=suppB]
\sname{Supplement B}\label{suppB}
\stitle{Italy, Death rates}
\slink[doi]{10.1214/10-AOAS394SUPPB}
\slink[url]{http://lib.stat.cmu.edu/aoas/394/supplementB.txt}
\sdatatype{.txt}
\sdescription{Italian population death rates. The data are downloaded
from the Human Mortality database and are indexed by calendar year
during the period 1950--2005. They are divided by sex and by single year
of age for ages from 0 to 100. For each gender and for each calendar
year, the death rates are given by the ratio between the ``Number of
deaths'' and the ``Exposure to risk.''}
\end{supplement}
%


\printaddresses


\begin{thebibliography}{26}

\bibitem[\protect\citeauthoryear{}{2002}]{BMS}
\textsc{Booth, H., Maindonald, J.} and \textsc{Smith, L.} (2002).
Applying Lee--Carter under conditions of variable mortality decline.
\textit{Population Studies} \textbf{56} 325--336.

\bibitem[\protect\citeauthoryear{}{1976}]{BJ}
\textsc{Box, G. E. P} and \textsc{Jenkins, G. M.} (1976).
\textit{Time Series Analysis for Forecasting and Control}.
Holden-Day, San Francisco.
\MR{0436499}

\bibitem[\protect\citeauthoryear{}{2002}]{BDV}
\textsc{Brouhns, N., Denuit, M.} and \textsc{Vermunt, J. K.} (2002).
A Poisson log-bilinear regression approach to the construction of
projected life tables. \textit{Insurance Math. Econom.}
\textbf{31} 373--393.
\MR{1945540}

\bibitem[\protect\citeauthoryear{}{2009}]{CBDCE}
\textsc{Cairns, A., Blake, D., Dowd, K., Coughlan, G. D., Epstein, D., Ong,
A., Balevich~I., Brouhns, N., Denuit, M.} and \textsc{Vermunt, J. K.} (2009).
A quantitative comparison of stochasting mortality models using data
from England and
Wales and the United States. \textit{N. Amer. Actuar. J.}
\textbf{13} 1--35.

\bibitem[\protect\citeauthoryear{}{2006}]{CMI}
CMI (2006). Stochastic projections methodologies:
Further progress and P-spline model feature, example results and
implications. Working Paper 20, Continuous Mortality
Investigation.

\bibitem[\protect\citeauthoryear{}{2004a}]{CDEa}
\textsc{Currie, I. D., Durban, M.} and \textsc{Eilers, P. H. C.} (2004a).
Smoothing and forecasting mortality rates. \textit{Statist.
Model.} \textbf{4} 279--298.
\MR{2086492}

\bibitem[\protect\citeauthoryear{}{2004b}]{CDEb}
\textsc{Currie, I. D., Durban, M.} and \textsc{Eilers, P. H. C.} (2004b).
Generalized linear array models with applications to multidimensional
smoothing. \textit{J. Roy. Statist. Soc. Ser. B}
\textbf{68} 259--280.
\MR{2188985}

\bibitem[\protect\citeauthoryear{}{2008}]{C}
\textsc{Currie, I. D.}, (2008).
Smoothing overparameterized
regression models. In \textit{Proceedings of 23rd International Workshop
on Statistical Modelling}, Utrecht, 194--199.


\bibitem[\protect\citeauthoryear{}{2007}]{DDE}
\textsc{Delwarde, A., Denuit, M.} and \textsc{Eilers, P.} (2007).
Smoothing the Lee--Carter and Poisson log-bilinear models for mortality
forecasting. \textit{Statist. Model.} \textbf{7} 29--48.

\bibitem[\protect\citeauthoryear{}{2010}]{DBCEM}
\textsc{Dowd, K., Blake, D., Cairns,
A. J. G., Eilers, P. H. C.} and \textsc{Marx, B. D.} (2010).
Facing up to uncertain life
expectancy: The longevity fan charts. \textit{Demography} \textbf{47}
67--78.

\bibitem[\protect\citeauthoryear{}{1993}]{ET}
\textsc{Efron, B.} and \textsc{Tibshirani, R. J.} (1993).
\textit{An Introduction to the Bootstrap}. Chapman and Hall, London.
\MR{1270903}

\bibitem[\protect\citeauthoryear{}{1996}]{EM}
\textsc{Eilers, P. H. C.} and \textsc{Marx, B. D.} (1996).
Flexible smoothing with B-splines and penalties. \textit{Statist.
Sci.} \textbf{11} 89--121.
\MR{1435485}



\bibitem[\protect\citeauthoryear{}{2003}]{RHb}
\textsc{Haberman, S.} and \textsc{Renshaw, A. E.}  (2003).
On the forecasting of mortality reduction factors. \textit{Insurance
Math. Econom.} \textbf{32} 379--401.

\bibitem[\protect\citeauthoryear{}{2008}]{RH}
\textsc{Haberman, S.} and \textsc{Renshaw, A. E.}  (2008).
On simulation-based approaches to risk measurement in mortality with
specific reference to Poisson Lee--Carter modelling.
\textit{Insurance Math. Econom.} \textbf{42} 797--816.


\bibitem[\protect\citeauthoryear{}{1994}]{H}
\textsc{Hamilton, J. D.} (1994).
\textit{Time Series Analysis}. Princeton Univ. Press, Princeton, NJ.
\MR{1278033}


\bibitem[\protect\citeauthoryear{}{2007}]{Hu}
\textsc{Hyndman, R. J.} and \textsc{Ullah, S.} (2007).
Robust forecasting of mortality and fertility rates: A~functional data
approach. \textit{Comput. Statist. Data Anal.} \textbf{51} 4942--4956.

\bibitem[\protect\citeauthoryear{}{2006}]{KSH}
\textsc{Koissi, M. C., Shapiro, A. F.} and \textsc{Hognas, G.} (2006).
Evaluating and extending the Lee--Carter model for mortality
forecasting: Bootstrap confidence interval. \textit{Insurance
Math. Econom.} \textbf{26} 1--20.
\MR{2197300}

\bibitem[\protect\citeauthoryear{}{1992}]{LC}
\textsc{Lee, R. D.} and \textsc{Carter, L. R.} (1992).
Modelling and forecasting US mortality. \textit{J.
Amer. Statist. Assoc.} \textbf{87} 659--671.

\bibitem[\protect\citeauthoryear{}{2001}]{LM}
\textsc{Lee R. D.} and \textsc{Miller, T.} (2001).
Evaluating
the performance of the Lee--Carter method for forecasting mortality.
\textit{Demography} \textbf{38} 537--549.



\bibitem[\protect\citeauthoryear{}{2003a}]{RHa}
\textsc{Renshaw, A. E.} and \textsc{Haberman, S.} (2003a).
Lee--Carter mortality forecasting: A parallel generalised linear
modelling approach for England and Wales mortality
projections.
\textit{Appl. Statist.} \textbf{52} 119--137.
\MR{1959085}


\bibitem[\protect\citeauthoryear{}{2003b}]{RHc}
\textsc{Renshaw, A. E.} and \textsc{Haberman, S.} (2003b).
Lee--Carter mortality forecasting with age specific enhancement.
\textit{Insurance Math. Econom.} \textbf{33} 255--272.
\MR{2039286}\


\bibitem[\protect\citeauthoryear{}{2010}]{WL}
\textsc{Wang, C.} and \textsc{Liu, Y.} (2010).
Comparison of mortality modelling and forecasting---empirical evidence
from Taiwan. Lee--Carter mortality forecasting with age specific
enhancement. \textit{Int. Res. J.~Fin. Econ.} \textbf{37} 46--55.
\end{thebibliography}
\end{document}